\documentclass[preprint,aps,pra,floatfix]{revtex4-1}
\usepackage{amsmath}
\usepackage{amssymb}
\usepackage[pdftex]{graphicx}
\newcommand{\mrm}{\mathrm}
\newcommand{\ket}[1]{{\left| {#1} \right\rangle}}

\newcommand{\avg}[1]{\left\langle {#1} \right\rangle }

\begin{document}

\title[Displacement operators: the classical face of their quantum phase]{Displacement operators: the classical face of their quantum phase}

\author{Amar C. Vutha}
\affiliation{Department of Physics, University of Toronto, 60 St. George Street, Toronto ON M5S 1A7, Canada}
\email{vutha@physics.utoronto.ca}
\author{Eliot A. Bohr}
\affiliation{Department of Physics \& Astronomy, University of California Los Angeles, Los Angeles, CA 90095, USA}
\author{Anthony Ransford}
\affiliation{Department of Physics \& Astronomy, University of California Los Angeles, Los Angeles, CA 90095, USA}
\author{Wesley C. Campbell}
\affiliation{Department of Physics \& Astronomy, University of California Los Angeles, Los Angeles, CA 90095, USA}
\author{Paul Hamilton}
\affiliation{Department of Physics \& Astronomy, University of California Los Angeles, Los Angeles, CA 90095, USA}

\begin{abstract}
In quantum mechanics, the operator representing the displacement of a system in position or momentum is always accompanied by a path-dependent phase factor. In particular, two non-parallel displacements in phase space do not compose together in a simple way, and the order of these displacements leads to different displacement composition phase factors. These phase factors are often attributed to the nonzero commutator between quantum position and momentum operators, but such a mathematical explanation might be unsatisfactory to students who are after more physical insight. We present a couple of simple demonstrations, using classical wave mechanics and classical particle mechanics, that provide some physical intuition for the phase associated with displacement operators.
\end{abstract}

\maketitle
\section{Introduction}\label{sec:Introduction}
The displacement operator in quantum mechanics effects displacements in phase space: it changes the position or the momentum (or both) of a system \footnote{See, for example, \cite{Goldstein2001} for an introduction to phase space.}. This operator is usually introduced to students in the context of the quantum harmonic oscillator (see e.g., \cite{Sakurai2014}). The displacement operator is particularly useful in quantum optics, where it enables some convenient methods to describe quantum states of the electromagnetic field (see e.g., \cite{Gerry2005}). In this context, the displacement operator was introduced by Glauber \cite{Glauber1963} in the form
\begin{equation}\label{eq:def_displacement}
\mathcal{D}(\alpha) = e^{\alpha a^\dagger - \alpha^* a},
\end{equation}
where $a^{\dagger}$ and $a$ are respectively the raising (photon
creation) and lowering (photon annihilation) operators of the harmonic
oscillator (electromagnetic field), and $\alpha$ is a complex number. The operator $\mathcal{D}(\alpha)$ displaces the position or momentum of the harmonic oscillator (or both), depending upon the complex number $\alpha$. Real values of $\alpha$ correspond to pure position displacements, while imaginary values of $\alpha$ correspond to pure momentum displacements.

In particular, displacement operators provide a natural way to define \emph{coherent states}, quantum states in which the average position $\avg{x(t)}$ and momentum $\avg{p(t)}$ oscillate sinusoidally as they would in a classical oscillator (unlike energy eigenstates of a harmonic oscillator, wherein $\avg{x}=\avg{p}=0$) \cite{Schrodinger1926,Gerry2005,Glauber1963,Sakurai2014}. A coherent state $\ket{\alpha}$ is the result of a displacement operator $\mathcal{D}(\alpha)$ applied to the ground state of the harmonic oscillator $\ket{0}$,
%
%
%
%
%
%
\begin{equation}
\ket{\alpha} \equiv \mathcal{D}(\alpha) \ket{0}.
\end{equation}

Unsurprisingly, the effect of a displacement operator can be undone by another operator that implements the opposite displacement, $\mathcal{D}(\alpha)^{-1} = \mathcal{D}(-\alpha)$. However, a peculiar feature of displacement operators is that two displacement operators do not always combine in a simple form: the operator for a phase-space displacement $\mathcal{D}(\alpha + \beta)$ is in general not equal to the product of the two displacement operators $\mathcal{D}(\alpha)$ and $\mathcal{D}(\beta)$. Instead, we have
\begin{equation}\label{eq:Composition_Rule}
\mathcal{D}(\alpha + \beta) = \mathcal{D}(\beta) \mathcal{D}(\alpha) \, \mathrm{e}^{\left(\alpha \beta^* - \alpha^* \beta \right)/2},
\end{equation}
which can be shown by applying the Baker-Campbell-Hausdorff theorem to Eq.~(\ref{eq:def_displacement}), and using the commutation relation $[a,a^\dagger] = 1$ \cite{Sakurai2014}.

The term $e^{\left(\alpha \beta^* - \alpha^* \beta \right)/2}$ is the exponential of an imaginary number and therefore represents a pure phase shift. We will refer to this as the \emph{displacement composition phase} (DCP), the physical interpretation of which is our central focus in this paper.

It is sometimes claimed that the DCP is physically irrelevant (cf. \cite{Gerry2005}), and should be disregarded as unmeasurable. This is only true if the DCP is global to the quantum state of a system. We emphasize the well-established fact that by applying displacements to only parts of a system's wavefunction, the DCP can indeed become measurable (see e.g., \cite{Luis2001}). For instance, the DCP will influence measurement outcomes in the interference between an arbitrary state $\ket{\gamma}$, and the state
\begin{equation}\label{eq:phase}
\ket{\gamma'} = \mathcal{D}(-\beta) \mathcal{D}(-\alpha)
\mathcal{D}(\beta) \mathcal{D}(\alpha) \ket{\gamma} =
\mathrm{e}^{\left(\alpha^* \beta - \beta^* \alpha \right)} \ket{\gamma}.
\end{equation}
We will use Eq.~(\ref{eq:phase}) to define the phase $\phi_{\mathrm{c}}$ via $\mathrm{i} \phi_{\mathrm{c}}\equiv \alpha^* \beta - \beta^* \alpha$. DCPs are routinely measured in experiments, including those using charged atoms confined in ion traps where these phase factors are central to the ``geometric phase gates'' used in quantum information processing \cite{Luis2001,Leibfried2003,PattyLeeJOB05,Nielsen2011,MizrahiAPB14}.
We emphasize that, since this phase factor arises from a displacement in general, the DCP is not just a feature of harmonic oscillator physics.

In this article, we explore the physical meaning of the DCP, and point out connections to classical mechanics that shed light on the physical mechanisms that are responsible for it. Since classical mechanics distinguishes between waves and particles, we describe connections to both wave and particle mechanics in order to develop an understanding of the quantum mechanical DCP. 

\section{Displacement operations in classical wave mechanics}\label{DOICWM}
We begin by showing how the DCP arises in classical wave mechanics. We start by asking how we can construct an operator that displaces a function, $f(x,t)$, by a distance $X$ in position \footnote{We will follow the convention of using upper-case symbols (such as $X$, $K$, and $P$) for constants and lower-case symbols (e.g. $x$, $k$ and $p$) for variables and operators.}.  Physically, $f(x,t)$ may represent a wave excitation of some medium, and for this discussion we will fix $t$ at some instant in time ($T_0$) and remove the time-dependence of the function to confine the discussion to the spatial domain \footnote{The opposite approach (fixing $x$ and discussing $\omega$ and $t$ displacements) is of course almost identical and leads to the same conclusions.}. We follow the standard procedure of considering an infinitesimal position displacement $D_{\mathrm{p}}.(\epsilon X) = 1 + \epsilon X k$ (where we take $\epsilon \ll 1$) and seeking $k$, the generator of infinitesimal position displacements:
\begin{eqnarray}
D_{\mathrm{p}}(\epsilon X)f(x) & = & f(x - \epsilon X)
\nonumber \\
& = & f(x) -\epsilon X\; \frac{\mathrm{d}}{\mathrm{d}x}f +
\mathcal{O}(\epsilon^2).
\end{eqnarray}
From this, we readily identify $k = -\frac{\mathrm{d}}{\mathrm{d}x}$. Extending this to finite displacements gives us
\begin{equation}
D_{\mathrm{p}}(X) = \mathrm{e}^{-X \frac{\mathrm{d}}{\mathrm{d}x}}.
\end{equation}
Displacements in spatial frequency, which translate the Fourier transform of $f(x)$ in the Fourier-conjugate ``$k$-space,'' are produced by imposing a phase shift that is linear in position,
\begin{equation}
D_{\mathrm{f}}(K) = \mathrm{e}^{\mathrm{i} K x},
\end{equation}
a standard result of Fourier transforms \cite{MorseAndFeshbach}, with the simple physical interpretation as the extra phase factor picked up by a wave that has to traverse an extra distance.

\begin{figure}
\begin{center}
\includegraphics[width=0.6\columnwidth]{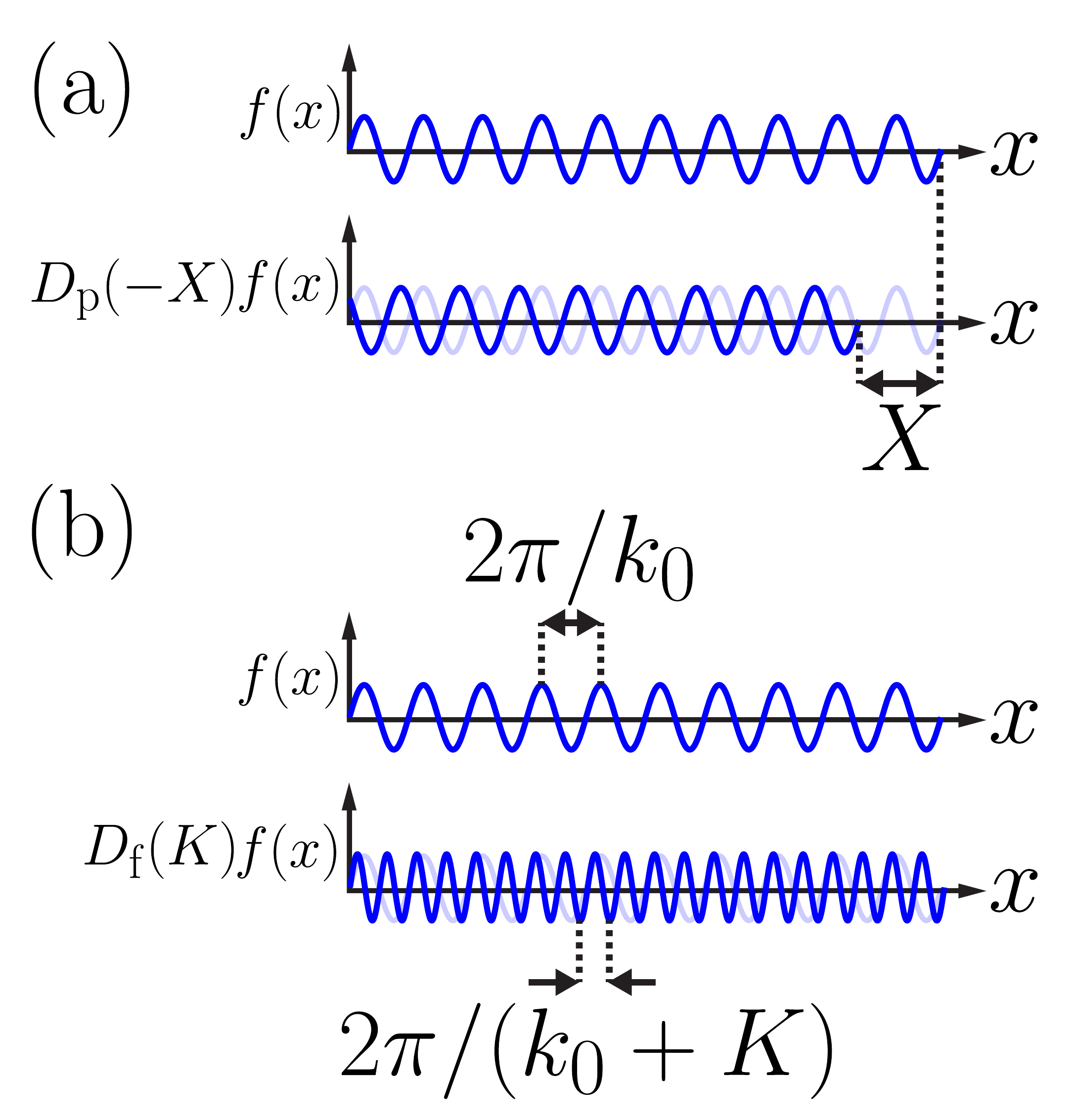}
\caption{The effect of (a) a position displacement operation or (b) a spatial-frequency displacement operation is shown on a sine function.}
\label{fig:sine_wave_cartoon}
\end{center}
\end{figure}

Figure \ref{fig:sine_wave_cartoon} shows the effect of $D_{\mathrm{p}}(-X)$ and $D_{\mathrm{f}}(K)$ on $f(x)=\sin(k_0 x)$. The choice of a negative sign for the position displacement example is motivated by its experimental implementation as a spatial delay in \S\ref{subsec:experiment}. We note that it can be seen in Fig.~\ref{fig:sine_wave_cartoon}(b) that the phase of $f(x)$ at $x=0$ is unchanged by the spatial-frequency displacement.

If we consider a case where $f(x)$ is well-behaved and is either periodic or restricted to a finite region of space of length $L$, we can write it as a Fourier series:
\begin{equation}
f(x) = \sum_{n=-\infty}^{\infty}c_n \mathrm{e}^{\mathrm{i}2 \pi n x/L}.
\end{equation}
Linear operations on $f(x)$ can then be computed individually on the elements of this series, and we find after some simple algebra that
\begin{equation}
D_{\mathrm{f}}(-K)D_{\mathrm{p}}(-X)D_{\mathrm{f}}(K)D_{\mathrm{p}}(X) \, f(x) = \mathrm{e}^{\mathrm{i}
  \phi_{\mathrm{c}}} \, f(x),
\end{equation}
where $\phi_{\mathrm{c}} = XK$.

The interpretation this suggests for the appearance of the DCP $\phi_{\mathrm{c}}$ is that the wave oscillates a different number of times when traversing the extra distance introduced by the second displacement than it would have for the distance removed by the first. This effect can be traced directly to the fact that the wave's spatial frequency has been changed between the two.  To put this another way, as one can see in Fig.~\ref{fig:sine_wave_cartoon}, the phase remains unchanged only at the origin when a frequency shift is applied; changing the ordering of position and frequency shifts therefore generally leads to a phase shift between the two possibilities.

We can use this result to address our original question, \emph{where does the DCP of the quantum harmonic oscillator (Eq.~\ref{eq:Composition_Rule}) come from?}  It follows from the choice of $\alpha \equiv X/x_0$ and $\beta \equiv \mathrm{i} P/p_0$ with $P = \hbar K$, which gives the same result as our classical expression: $\phi_{\mathrm{c}} = XP/\hbar$. These choices are well-motivated by the fact that the real and imaginary parts of the coherent state index correspond to the position and momentum displacements of the quantum harmonic oscillator ground state.
%
%
Since the DCP is a property of the displacements themselves (as opposed to the state/function on which they operate), there is no need to identify a definite frequency for the original wave.  This illustrates the fact that this classical DCP is not merely analogous to the quantum version, but is in fact the same for any situation where the displacement in frequency space, $K$, can be unambiguously identified.


The extension of these ideas to temporal displacements by $T$ and temporal-frequency displacements by $\Omega$ follows an identical development, yielding $\phi_{\mathrm{c}} = \Omega T$.  Following a similar interpretation as that above, for the time-domain process $\phi_{\mathrm{c}} = \Omega T$ is just the phase difference introduced between two waves with frequencies $\Omega_0$ and $\Omega_0 + \Omega$ when they are delayed by a duration $T$.

\subsection{Experimental demonstration of the DCP in a classical system}\label{EDOTDOPIACS}\label{subsec:experiment} 
As an illustration of the physical relevance of the DCP $\phi_{\mathrm{c}}$, we present a simple experiment for measuring it in a classical system.  We stress that this experiment is essentially pedagogical in nature, and is intended to demonstrate the basic physics that leads to the appearance of the DCP.  Since our system involves a laser beam, we also wish to make it clear that this experiment performs displacements in the classical wave phase space defined by position and spatial frequency, and does not involve the field quadrature phase space that may be familiar from quantum optics (cf. \cite{Gerry2005}).

\begin{figure}
\begin{center}
\includegraphics[width=0.6\columnwidth]{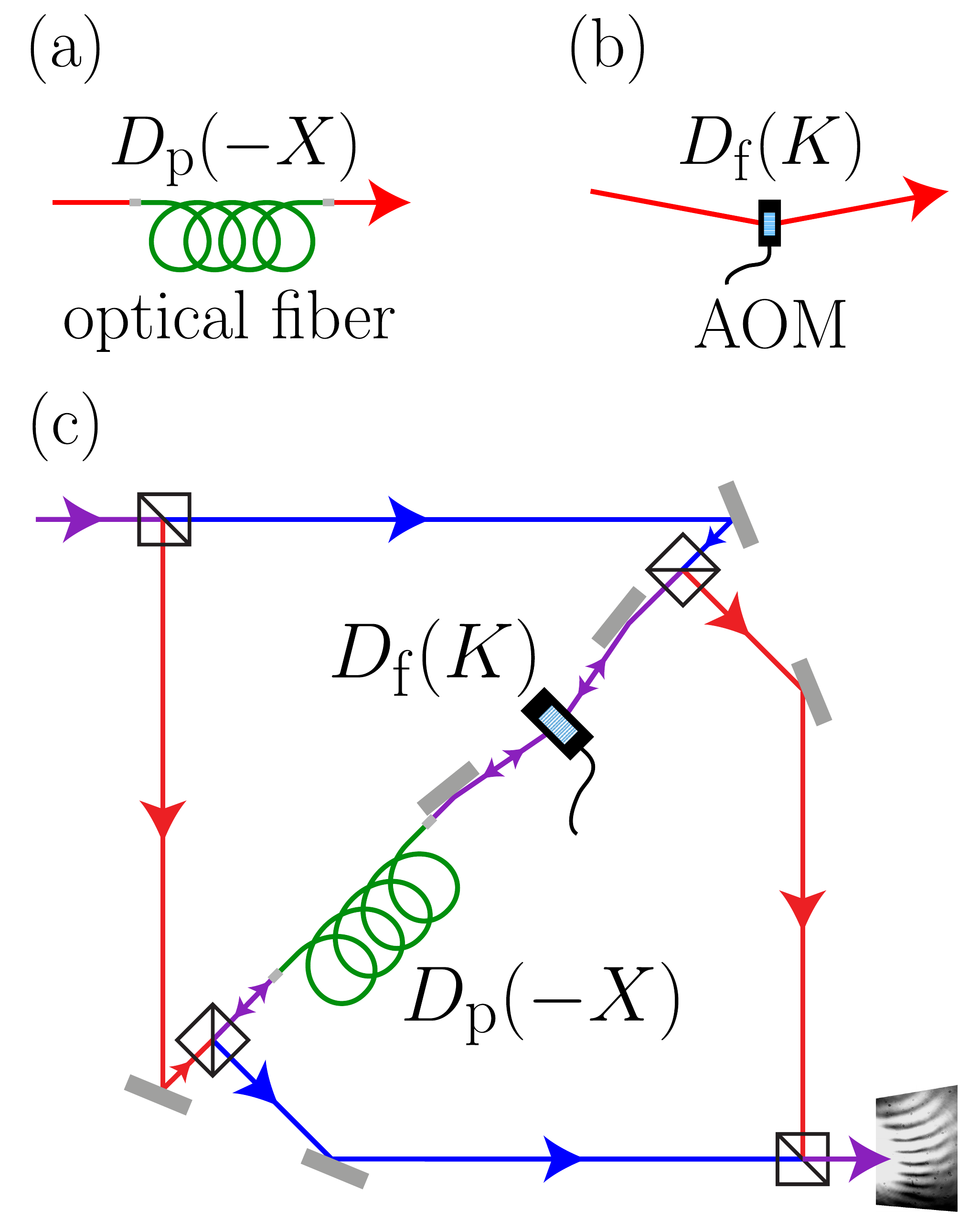}
\caption{Phase-space displacement operations on a laser beam showing the effect of (a) displacement in position by insertion of an optical fiber delay line and (b) displacement in spatial frequency implemented as a wavelength shift from Bragg reflection in an acousto-optic modulator (AOM). A diagram of the interferometer for measuring $\phi_{\mathrm{c}}$ is shown in (c), where the introduction of a slight relative tilt between the two output beams generates a static spatial fringe pattern, shown at the bottom right.  A scan of the AOM frequency results in a spatial shift of the fringe pattern's phase, which is read by a digital camera.  One arm of the interferometer is shown in red and the other in blue, while portions where the two beams are overlapped are shown in purple.  All beam splitters are 50/50 and non-polarizing, and light that is rejected from unused beam splitter and AOM ports has been omitted for clarity.}
\label{fig:laser_show}
\end{center}
\end{figure}

The apparatus consists of an optical interferometer where a laser beam is spatial-frequency shifted and spatially displaced by the same amounts in each interferometer arm, but in opposite orders, as shown in Fig.~\ref{fig:laser_show}.  Following Section \ref{DOICWM}, the spatial-domain picture can be isolated from any time-dependence by describing the state of the system at any fixed moment in time $t=T_0$. This particular interferometer does not directly compare an undisplaced system to one that has been displaced around a closed loop (such as Eq.~(\ref{eq:phase})), but instead encloses a loop by displacing a system around two different trajectories that start and end in the same location in phase space (analogous to the paths $A \!\!\rightarrow \!\!B \!\!\rightarrow \!\!C$ and $A\!\!\rightarrow \!\!D\! \rightarrow \! C$ in Fig.~(\ref{fig:phase_space_loop}) with the vertical axis denoting spatial frequency $k$ instead of $p$). The reason for this is simply that the latter was more experimentally convenient to implement, and the two protocols will result in precisely the same value for $\phi_{\mathrm{c}}$, a result that follows from the description of $\phi_{\mathrm{c}}$ as the area enclosed in phase space.

Position displacements are performed by insertion of an optical delay line in the form of an optical fiber, as shown in Fig.~\ref{fig:laser_show}(a).  The magnitude of the actual displacement in this case requires that we take into account the correct phase velocity of the wave in the fiber when calculating the optical path length.  Displacements of the spatial frequency are performed by sending the laser beam through a moving diffraction grating in an acousto-optic modulator (AOM), which is driven at a temporal frequency  $\Omega_{\mathrm{rf}}$. (Interested readers are referred to e.g.,~\cite{SalehTeich} for an introduction to acousto-optics.) This shifts the optical spatial frequency of the laser beam by $K = \Omega_{\mathrm{rf}}/c$, as depicted in Fig.~\ref{fig:laser_show}(b).  This operation differs from the ideal $D_{\mathrm{f}}(K)$ in that it adds the phase of the radio frequency signal ($\Omega_{\mathrm{rf}}$) going to the AOM to the optical beam.  However, if this phase can be made identical in both arms of the interferometer (Fig.~\ref{fig:laser_show}(c)), it will not contribute to the phase shift measured in the experiment, and we can therefore set it equal to zero in our calculations.

\begin{figure}
\begin{center}
\includegraphics[width=0.6\columnwidth]{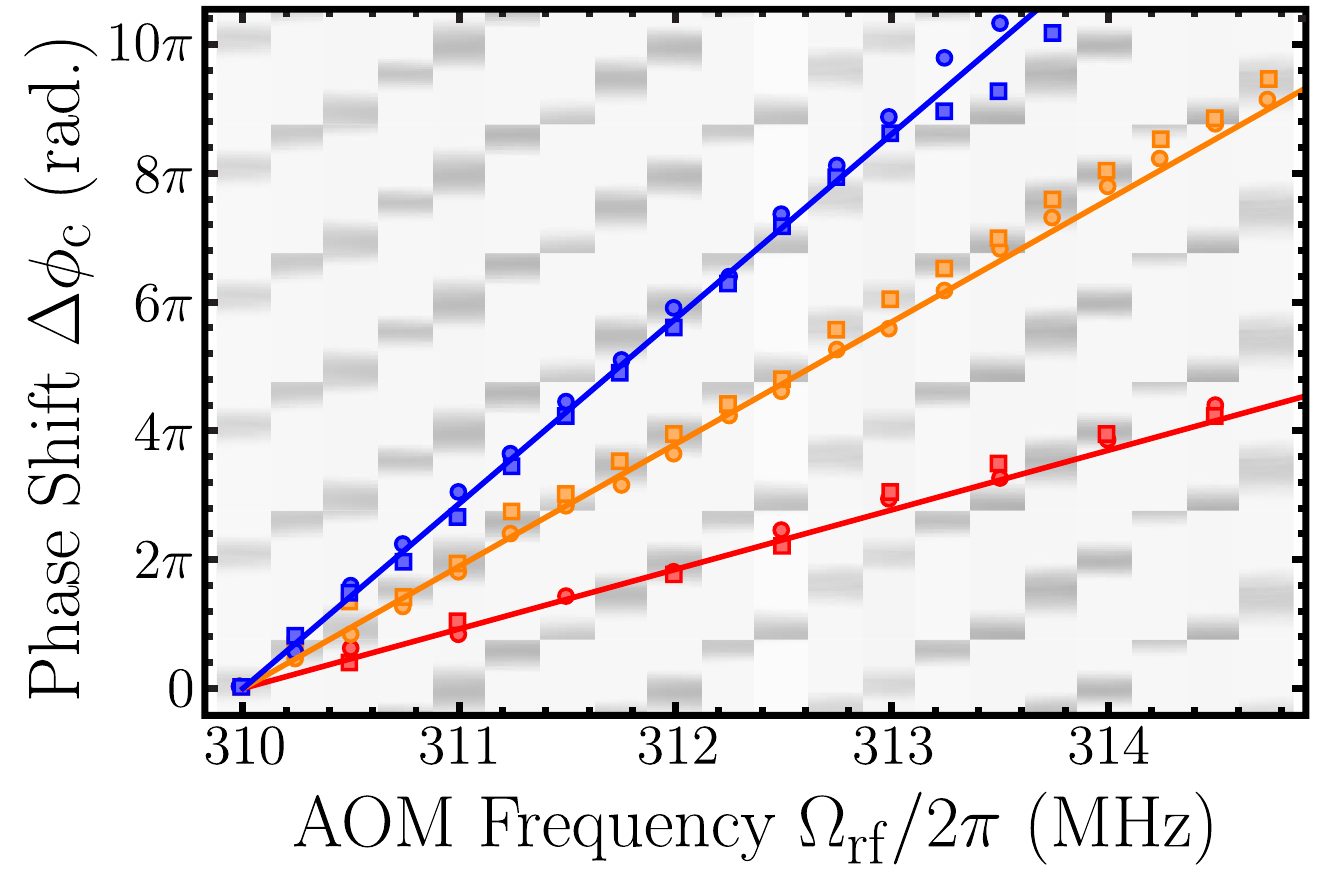}
\caption{Measurement of the displacement composition phase in a
  table-top system.  The measured phase shift
  ($\Delta\phi_{\mathrm{c}}$) is plotted for three different lengths
  of optical fiber: $95 \mbox{ m}$ (red), $195 \mbox{ m}$ (orange),
  and $295 \mbox{ m}$ (blue).  Points taken with the frequency scan
  moving upward (circles) and downward (squares) are shown to reflect
  the impact of possible fringe drifts during the measurements.  The
  background of the plot is a composite image made from camera images
  of the spatial fringe pattern at each AOM setting for the $X = 195
  \mbox{ m}$ fiber.  The solid curves are the theoretical predictions
  (as opposed to fits) using $\Delta \phi = \Delta K \,X$, where $X$
  is the manufacturer-specified length of the optical fiber times its effective refractive
  index.}
\label{fig:data_plot}
\end{center}
\end{figure}

We implement the interferometer in such a way that the same fiber and AOM are used simultaneously by each arm of the interferometer, shown schematically in Fig.~\ref{fig:laser_show}(c).  This geometry stabilizes the fringes against a few sources of technical noise (such as fiber vibration and temperature changes) and, as discussed above, ensures that the phase of the rf signal going to the AOM is identical for the two arms.  We introduce a slight tilt between the two beams combining on the last beam splitter to create a static, spatial fringe pattern that is recorded by a digital video camera (shown in the background of Fig.~\ref{fig:data_plot}).  The fringes are then fit to a (spatial) sinusoid to extract their phase.

To measure $\phi_{\mathrm{c}}$, we perform a differential measurement
comparing the phase of the interference (the spatial position of the
static fringes recorded by the camera) for one value of the AOM
frequency to the phase after a known frequency shift of $\Delta
\Omega_{\mathrm{rf}} \equiv \Delta K c$ has been applied.  If the
DCP is indeed given by $\phi_\mathrm{c} = K\, X$, the
slope of the interference phase vs.\ AOM frequency shift will yield
$\Delta \phi_{\mathrm{c}}/\Delta K = X$, which is dominated by the
length of the fiber delay.  This differential measurement (that is,
measuring the phase change as the frequency is changed) allows us to
confirm the form of $\phi_{\mathrm{c}}$ without requiring knowledge of
the absolute optical path length of the delay ($X\sim 100 \mbox{ m}$),
which would otherwise need to be known to within a fraction of a
wavelength of light.

Example measurements for three different spatial displacements (fiber
lengths) are shown in Fig.~(\ref{fig:data_plot}).  The solid lines are
theory based on the manufacturer-specified lengths and effective phase
velocities in the fibers.  As expected, we find excellent agreement
between the theoretical value predicted by quantum mechanics, that
predicted by classical mechanics (which is identical, as shown in
Section \ref{DOICWM}), and the measured phase shift in this classical
wave system.  At no point was knowledge of the absolute momentum or
energy of single photons required, nor was a description of the system
in terms of photons necessary.

While the results of these measurements are not surprising, they
serve to illustrate how phase factors from compositions of phase-space
displacements arise -- the frequency-shifted classical wave simply
oscillates a different number of times in the delay line than the wave
that traversed the delay line before it got shifted. This results in
a phase difference between the two that is equal to the product
of the optical path length difference and the spatial frequency shift. This
connection may aid in understanding the reason this DCP
arises in quantum systems by framing it in the context of an
intuitive, accessible table-top experiment. 

\section{Displacement operations in classical particle mechanics}\label{sec:DOICPM}

Having provided an intuitive picture for the DCP with classical waves,
we move on to examine classical particle mechanics in order to provide another 
simple picture for the DCP. We first review how classical canonical transformations produce displacements, and then demonstrate how these lead to a connection between classical phase space and the DCP.

We will describe the motion of a particle at a point $(X_0,P_0)$ in phase space. Let us construct a canonical transformation that displaces this point to a new location $(X_0+X,P_0 + P)$, by working backward from the desired
equations of motion. This canonical transformation is equivalent to
constructing a Hamiltonian function $H$ that generates the desired
transformations $X_0 \to X_0 + X$ and $P_0 \to P_0 + P$ as a consequence of  Hamilton's equations of motion
\begin{equation}
\begin{split}
\frac{\mathrm{d}x}{\mathrm{d}t} & = \frac{\partial H}{\partial p} \\
\frac{\mathrm{d}p}{\mathrm{d}t} & = -\frac{\partial H}{\partial q}.
\end{split}
\end{equation}
Recall that these equations are a special case of the more general
statement that the rate of change of any phase space function
$F(x,p)$, along a trajectory generated by a Hamiltonian $H$, is
\begin{equation}
\frac{\mathrm{d}F}{\mathrm{d}t} = \{F,H\}.
\end{equation}
Here $\{F,H\}$ denotes the classical Poisson bracket of the functions
$F$ and $H$.

The first-order, linear differential equations that generate a
displacement by an amount $(X,P)$ in a duration $T$ are
\begin{equation}
\begin{split}
\frac{\mathrm{d}x}{\mathrm{d}t} & = \frac{X}{T} \\
\frac{\mathrm{d}p}{\mathrm{d}t} & = \frac{P}{T}.
\end{split}
\end{equation}
We find, then, that displacement by $(X,P)$ can be generated by the static
Hamiltonian function
\begin{equation}\label{eq:displacement_hamiltonian}
H_\mrm{dis} = \frac{pX - xP}{T}
\end{equation}
in a time $T$, regardless of what the actual value of $T$ is.

As an aside, note that the operators $\hat{x} \equiv x_0 \left( a + a^\dagger\right)/2$, $\hat{p} \equiv p_0 \left( a -
a^\dagger\right)/(2 \mathrm{i} )$ and $\alpha = \frac{X}{x_0} + \mathrm{i} \frac{P}{p_0}$ inserted into $\hat{H}_\mrm{dis}$, the quantized version of Eq.(\ref{eq:displacement_hamiltonian}), immediately lead to Equation (\ref{eq:def_displacement}) for a harmonic oscillator with mass $m$ and resonant frequency $\omega_0$ (with $x_0 \equiv \sqrt{2\hbar/(m \omega_0)}$ and $p_0 \equiv \sqrt{2 \hbar m \omega_0}$).

Let us now figure out how to calculate the phase that arises in a quantum treatment of the same process. We observe from Eq.~(\ref{eq:phase}) that the DCP $\phi_{\mathrm{c}}$ appears when, for instance, a quantum state is
displaced around a parallelogram (with sides $\alpha, \beta$) on the
complex plane. We therefore begin by examining the classical
description of displacement of a particle around a rectangle in phase
space.

\begin{figure}[htp!]
\begin{center}
\includegraphics[width=0.6\columnwidth]{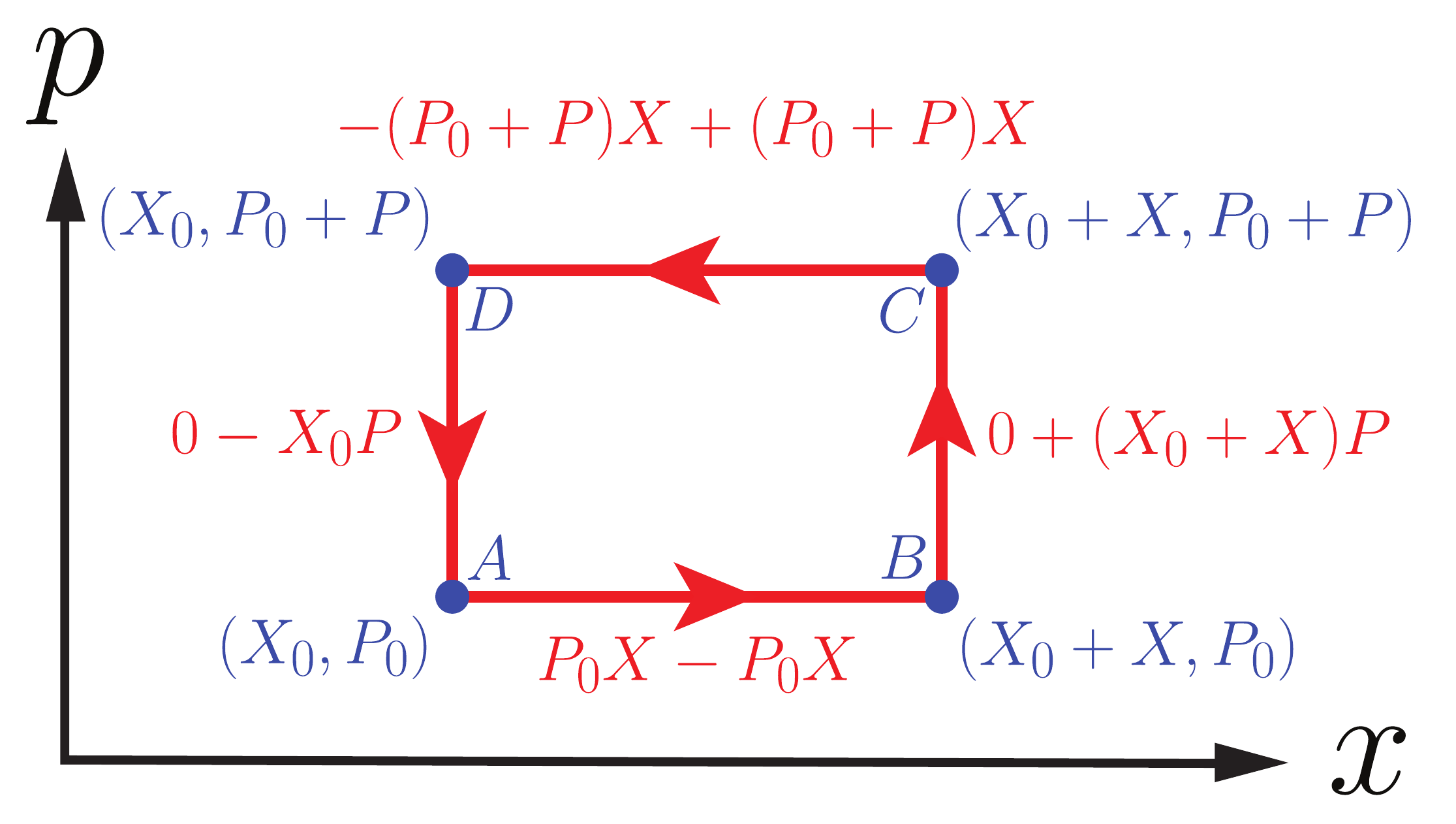}
\caption{A rectangle in phase space, beginning at the arbitrary point
  $(X_0,P_0)$. Shown in red are the contributions to the action
  integral, $S_i = \int p \, \mathrm{d}x - H_i \, \mathrm{d}t$, along
  each segment of the trajectory. The total action acquired in
  traveling around the rectangle in the positive sense
  (counter-clockwise) is $\sum S_i = XP= \hbar\, \phi_{\mathrm{c}}$.}
\label{fig:phase_space_loop}
\end{center}
\end{figure}

Consider a particle in phase space at the point $A = (X_0,P_0)$ that
is displaced as shown in Fig.~\ref{fig:phase_space_loop}: first,
along the $x$-axis to $B = (X_0 + X,P_0)$, then along the $p$-axis to
$C = (X_0 + X,P_0 + P)$, then to $D = (X_0, P_0 + P)$, and then back
to the original point $A = (X_0,P_0)$, so that the trajectory traces
out a rectangle $ABCD$. We will calculate two quantities for each
side of this rectangle:
\begin{enumerate}
\item[(i)] A Hamiltonian function $H$ that is a constant along the
  trajectory that is generated by it -- in other words,
  $\frac{\mathrm{d}H}{\mathrm{d}t} = \{H,H\} = 0$ -- which is just
  another way of saying that energy is conserved.
\item[(ii)] The phase factor picked up by a \emph{quantum} system that
  is transported along that phase space curve $\xi(t)$, given by $\mathrm{e}^{\mathrm{i}
    S/\hbar} = \mathrm{e}^{\frac{\mathrm{i}}{\hbar} \int L \mathrm{d}t}  = \mathrm{e}^{\frac{\mathrm{i}}{\hbar} \int p \, \mathrm{d}x - H \,
    \mathrm{d}t}$, where $S = S[\xi(t)]$ is the action
  integral along the classical path $\xi$.
\end{enumerate}

We can now write the contributions to the classical action
integral along the legs of this rectangle, shown in Fig.~\ref{fig:phase_space_loop}:
\begin{description}
\item[\textnormal{$A \to B$}] The Hamiltonian is $H_1 = pX/T_1$ , and the action integral is $S_1 = \int_A^B p \, \mathrm{d}x - H_1 \, \mathrm{d}t = P_0 X - P_0 X = 0$.
\item[\textnormal{$B \to C$}] The Hamiltonian is $H_2 = -xP/T_2$, and the action integral is $S_2 = \int_B^C p \, \mathrm{d}x - H_2 \, \mathrm{d}t = 0 + (X_0 + X) P$.
\item[\textnormal{$C \to D$}] The Hamiltonian is $H_3 = -pX/T_3$, and the action integral is $S_3 = \int_C^D p \, \mathrm{d}x - H_3 \, \mathrm{d}t = -(P_0 + P)X + (P_0 + P) X = 0$.
\item[\textnormal{$D \to A$}] The Hamiltonian is $H_4 = xP/T_4$, and the action integral is $S_4 = \int_D^A p \, \mathrm{d}x - H_4 \, \mathrm{d}t = 0 - X_0 P$.
\end{description}

Therefore any quantum state transported around a phase-space rectangle
$ABCD$ picks up a phase factor
\begin{equation}\label{eq:rectangle_action}
\mathrm{e}^{\mathrm{i}(S_1 + S_2 + S_3 + S_4)/\hbar} = \mathrm{e}^{\mathrm{i} XP/\hbar}.
\end{equation}
The acquired phase is equal to the geometrical area of the phase space rectangle (in units of $\hbar$). 

Having seen that the action integral due to displacement around any
phase space rectangle is equal to the area of the rectangle, reasoning
similar to the derivation of Stokes' theorem \cite{Hubbard2009} allows us to generalize this statement to the action $S(\mathcal{C})$ acquired by displacement around \emph{any} closed curve $\mathcal{C}$ in phase space:
$S(\mathcal{C}) = \int L \mathrm{d} t = \oint_\mathcal{C} p \,
\mathrm{d}x - H_\mrm{dis} \, \mathrm{d}t =
\mrm{Area}(\mathcal{C})$. This can also be shown graphically by
breaking up the area bounded by an arbitrary curve $\mathcal{C}$ into rectangular tiles, and observing that all the line integrals along interior edges have equal and opposite contributions from adjacent rectangles. Therefore, transporting a quantum state around a curve in phase space leads to a DCP that is always just the classically enclosed phase space area divided by $\hbar$.

This development of the DCP provides an immediate connection to concepts from classical mechanics in phase space. This result is also practically useful for  calculating phase shifts in matter-wave interference experiments with atoms and molecules \cite{Storey1994}.

\section{Summary}
We have illustrated some of the ways in which the form and behavior of the
quantum mechanical displacement operator are closely connected to
classical mechanics, with the goal of finding an intuitive picture for
the DCP. In particular, we have sought to understand the origins of the DCP that accompanies multiple non-parallel phase-space displacements. We have shown that this phase factor has a simple explanation in terms of classical wave mechanics, and that it can be readily connected to the classical action integral for particles. We hope that these demonstrations will be of interest to instructors of quantum mechanics and quantum optics, and to students who are beginning to explore topics such as geometric phase gates in quantum information physics.

\section*{acknowledgments}
We thank David Hayes and Eric Hudson for helpful discussions. ACV
acknowledges support from an NSERC Discovery Grant. WCC acknowledges
support from the National Science Foundation CAREER program under
award number 1455357. PH acknowledges support from the University of
California Office of the President's Research Catalyst Award
No. CA-16-377655. WCC and PH acknowledge support of the US Office
of Naval Research under award number N00014-17-1-2256.

\bibliography{displacement_operators}

\end{document}